\documentclass[11pt]{article}
\usepackage{graphicx}
\usepackage{amssymb}

\def\beqa{\begin{eqnarray}}
\def\eeqa{\end{eqnarray}}
\def\beq{\begin{equation}}
\def\eeq{\end{equation}}

\begin{document}

\title{Is there a  relationship between the mass of a SMBH and the kinetic energy
of its host elliptical galaxy?}

\author{A. Feoli\footnote{E-mail: feoli@unisannio.it} and D. Mele\\ Dipartimento di Ingegneria, Universit\`{a} del
            Sannio, \\Corso Garibaldi n. 107, Palazzo Bosco Lucarelli  \\ I-82100 - Benevento, Italy.}

 \maketitle

\begin{abstract}
We use a restricted sample of elliptical galaxies, whose
kinematical parameters inside the semimajor axis were calculated
correcting the effect of the integration of the light along the
line of sight, in order to analyze a possible relationship between
the mass of a Supermassive Black Hole (SMBH) and the kinetic
energy of random motions in the host galaxy. We find  $M_{BH}
\propto (M_{G} \sigma^{2})^\alpha$ with $0.87 \leq \alpha \leq 1$
depending on the different fitting methods and samples used.
 This result could be interpreted as a new fundamental
relationship or as a new way to explain the old $M_{BH} - \sigma$
law. In fact, the relations of the velocity dispersion both with
the mass of the SMBH ($M_{BH} \propto \sigma^{4.12}$) and with the
mass of the host galaxy ($M_{G} \propto \sigma^{2.16}$) induce us
to infer an almost direct proportionality: $M_{BH} \propto M_{G}
\sigma^{2}$. A similar relationship is found for the total kinetic
energy involving the rotation velocity too.
\end{abstract}

\section{Introduction}
The discovery in the nucleus of an increasing number of galaxies
of SMBHs and the estimation of their masses with several methods
have allowed the beginning of statistical analyses that try to
show the influence of a SMBH on the properties of  the host
galaxy. In particular, several relationships were proposed between
the mass of a central SMBH and the velocity dispersion \cite{fer},
\cite{geb}, \cite{tre}, the bulge luminosity or mass
 \cite{kor}\cite{mag} or the dark matter halo \cite{fer2},
 of the corresponding galaxy. Among them,
the relationship with the smallest scatter is $M_{BH} \propto
\sigma^{\alpha}$ where $3.75 < \alpha < 5.3$. The different values
of $\alpha$, found by several authors using different samples and
different fitting methods, are well discussed in the paper of
Tremaine et al. \cite{tre} that obtain $$Log M_{BH} = (4.02 \pm
0.32) Log (\sigma/200) + (8.13 \pm 0.06) \eqno(1)$$ where (and
also afterwards in the rest of our paper) $M_{BH}$ is expressed in
solar masses, $\sigma$ in $Km/s$ and logarithms are base 10.

 While the astronomers believe in
these relationships, in such a way that they use them for example
to predict the value of the mass of a SMBH in a galaxy of a
distant cluster \cite{mc}, from the theoretical point of view
there are many possible interpretations about the physical meaning
and the origin of these relations \cite{hae} \cite{bur}. In this
paper we verify the existence of a relationship between  the rest
energy of the SMBH and the kinetic energy of random motions inside
the semimajor axis of the host elliptical galaxy $M_{BH} c^2
\propto (M_G \sigma^{2})^\beta$. This could be a new fundamental
relation or it can be used (in a simple way if $\beta \simeq 1$)
to interpret and explain the $M_{BH} \propto \sigma^\alpha$ law
(where, for example, $\alpha \simeq 4$ as in eq. (1), but, in
principle, we could also find a different value), through a hidden
dependence of the mass of the galaxy on the velocity dispersion.
In that case a relation $M_G \propto \sigma^{\frac{\alpha}{\beta}
- 2}$ holds. Finally we consider the role of the rotation velocity
in a relationship between the rest energy of the SMBH and the
total kinetic energy.
 The paper is organized as
follows: in sect. 2 we describe the sample of galaxies chosen for
the statistical analysis and the model used to obtain the
kinematical data, in sect. 3 the results of the fitting procedure
are given and in sect. 4 the conclusions are drawn.

\section{The sample}
 As suggested by Gebhardt et al. \cite{geb}, a particular care has to
be devoted, in the choice of the sample, to find suitable
kinematical parameters that must be referred not only to the
central regions but for example to the effective radius, in order
to have a higher signal to noise ratio and to be less sensitive to
the direct influence of the gravitational field of the central
SMBH. In our case, we want to consider not only the velocity
dispersions but also the masses of the host galaxies. There are
several methods to compute the masses of galaxies, for example
using in different ways the Virial Theorem or starting from  the
Jeans equation or better with self consistent models and recently
using also  data on globular clusters, planetary nebulae and the
temperature and density distribution of the thermal X-ray gas. In
the past, one of us (A.F.) used two of these methods. When  the
mass of the whole galaxy was to be considered, the Virial Theorem
was used in its scalar form $2T + U = 0$ and,  knowing the
rotation velocity  and velocity dispersion, the corresponding
Virial Mass was derived \cite{bus}. On the contrary, if only the
part of the galaxy inside a radius $R_e$ is studied, the system is
not really isolated and the effect  of the matter at a radius
$r>R_e$ has to be considered as an unknown external pressure
\cite{cha}. This term and the rotation velocity  are often
neglected in the case of elliptical galaxies. In order to overcome
this problem, Busarello et al.\cite{busl} preferred to compute the
masses of a sample of 62 elliptical galaxies inside the effective
semimajor axis $A_e$ (related to the effective radius $R_e$ by the
formula $$ \frac{R_e}{A_e}= 1 - \frac{\epsilon}{2} \eqno(2)$$
where $\epsilon$ is the ellipticity), starting from the Jeans
equation describing the equilibrium of a spheroidally symmetric
system having eccentricity $e$ and an isotropic velocity
dispersion tensor: $$ \frac{1}{\rho
(r)}\frac{d}{dr}[\rho(r)\sigma^2(r)]-\frac{v^2(r)}{r} = -
\frac{4\pi G(1-e^2)^{1/2}}{r}\int^r_o \frac{dx \
x^2\rho(x)}{(r^2-x^2e^2)^{1/2}}\ \ , \eqno(3) $$ where $r$ is the
radius in the equatorial ($z=0$) plane, $\rho(r)$ is the (unknown)
spatial density, $\sigma(r)$ and $v(r)$ are the one--dimensional
velocity dispersion and rotation velocity respectively, and $G$ is
the constant of gravity \cite{bin}.

 The solution of Eq. (3) can be written in the form $\rho(r)= \rho_0 \times l(r)
$, where $l(r)$ is the luminosity density. Busarello et
al.\cite{blf} (hereafter BLF) assumed that the luminosity
distribution corresponds to the spatial deprojection of the
$r^{1/4}$ law. A simple analytical approximation for the
deprojection of the $r^{1/4}$ law has been derived by Mellier and
Mathez  \cite{mel}: $$l(r) = r^{-\beta} exp (-b r^{1/4})
\eqno(4)$$ where $\beta =0.855$ and $b=7.669$. Substituting this
solution together with $V(r)$ and $\sigma(r)$ in the equation (3)
an expression of $\rho_0$ as a function of $r$ can be obtained.
Computing the value of $\rho_0$ for each object in the sample at
10 different radii, the residuals with respect to its mean value
$<\rho_0>$ turn out to be very small (and will be used to estimate
the error $\Delta M$), and show no systematic trend with the
radius, thus supporting the hypothesis that $\rho_0=constant$ at
least in the inner regions $(r < A_e)$ \cite{busl}. So, the final
result for the mass density is $\rho(r)= <\rho_0> r^{-\beta} exp
(-b r^{1/4})$. Sometime it happens that two ways to estimate the
same quantity give different results; actually, in our case we
observe that this method to compute masses can lead to values that
differ even $30\%$ from  masses obtained using the Virial Theorem.

 We decide to adopt
along this paper all the kinematical parameters computed by
Busarello et al. and published for $\sigma$ and $V$ of 54
elliptical galaxies in BLF and for the mass and the specific
angular momentum in \cite{cur}. These results have the advantage
to be treated with the same method and the same fitting procedure;
furthermore they are all referred to the effective semimajor axis,
as we required, and are corrected for one projection effect. The
observable quantities are actually affected by two types of
projection effects: the inclination of the rotation axis with
respect to the line of sight and the integration of the light
along the line of sight. Busarello et al. \cite{blf} \cite{ber}
applied in fact a method to correct the second effect,
deprojecting the rotation velocity curves and the velocity
dispersion profiles. They start from a very simple model assuming
that an elliptical galaxy
\begin{enumerate}
\item has a spheroidal symmetry,
\item follows the de Vaucouleurs $r^{1/4}$ law whose spatial
deprojection was given by the approximated analytical expression
(4), \item the rotation axis is perpendicular to the line of
sight, \item the rotation velocity has cylindrical symmetry and
\item the velocity dispersion is isotropic and has spherical
symmetry. \end{enumerate} The method has been explained in details
in BLF where also the sources of kinematical data (except for
N3379 \cite{fra} and IC1459 \cite{fra2}) and the values of the
parameters used in the calculations are listed. It allows to
compute, starting from a fit of the experimental data, the
deprojected $V(r)$ and $\sigma(r)$ to be inserted in the equation
(3) in order to obtain the masses. From those two analytic
functions it is easy to obtain also the specific kinetic energies
due to rotation and random motions respectively and some others
kinematical parameters such as specific angular momentum, spin
etc.

 In this paper we consider as
the reference sample, the intersection between the set of
elliptical galaxies studied by Busarello et al. \cite{busl}
\cite{blf} \cite{cur} and the set of SMBH masses in Table 1 in the
paper of Tremaine et al. \cite{tre}. The masses and the
kinematical parameters of the resulting 14 galaxies are listed in
table 1 where $V$ is the luminosity weighted mean rotation
velocity inside $A_e$ (related to the rotational kinetic energy by
$T_V = M V^2/2$)
%such that $$ V^2 = \frac{\int V^2 (r) \rho d^3 x }{\int \rho d^3 x} $$
and  $$ \sigma^2 \equiv <\sigma^2_{yy}> = \frac{\int \sigma^2_{yy}
(r) \rho d^3 x }{\int \rho d^3 x} = \frac{2T_\sigma}{3M}
\eqno(5)$$ is the luminosity weighted mean of the line of sight
component, of the velocity dispersion tensor, assuming that the
mass-to-light ratio is constant inside $A_e$ and that the tensor
is isotropic ($T_\sigma$ is the corresponding kinetic energy).
\begin{figure}
  \centering
\resizebox{\hsize}{!}{\includegraphics{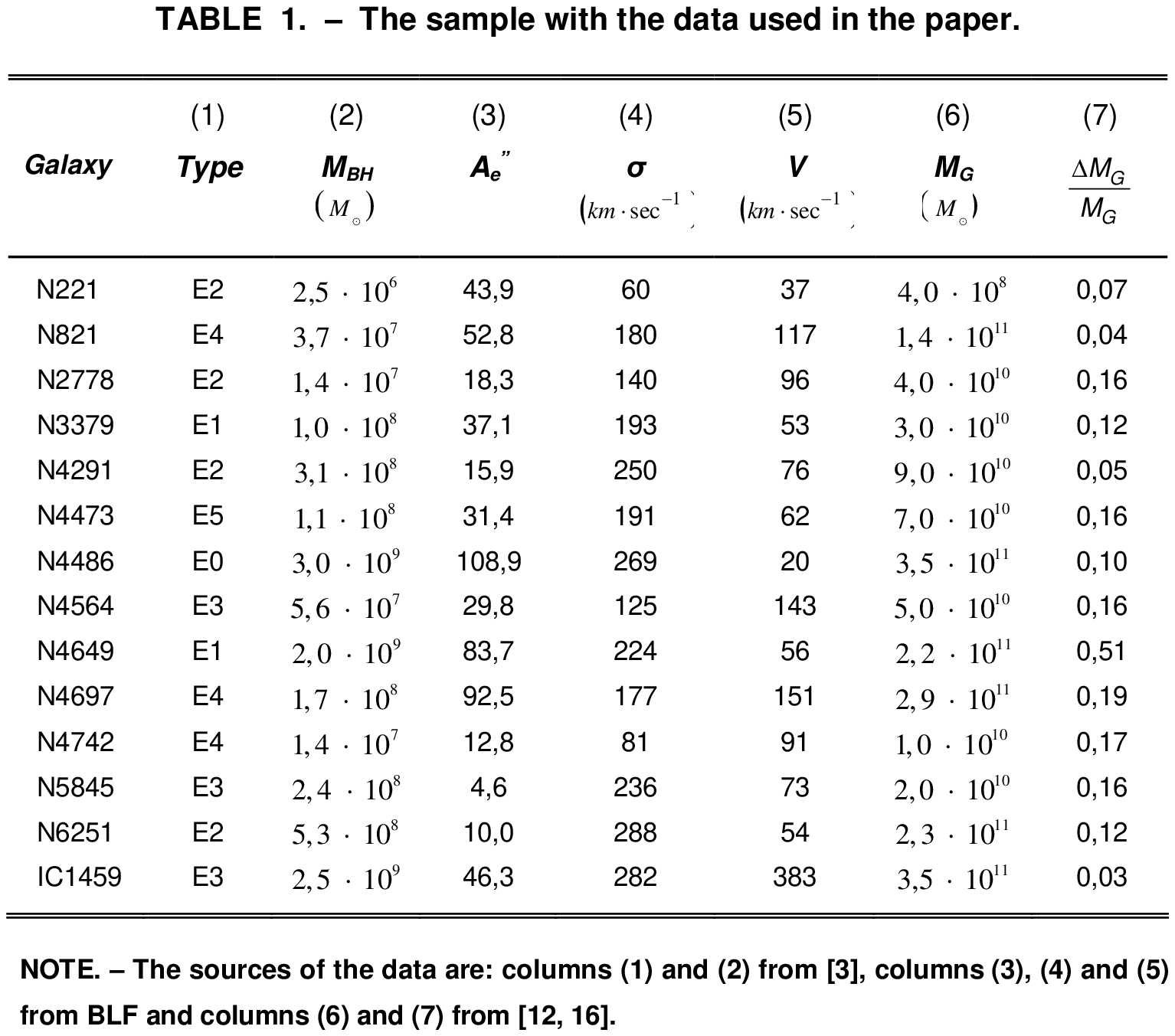}}
\end{figure}
 We do
not insert in our statistics the galaxy N221 because it has a mass
two orders of magnitude less than all the others and probably  a
different dynamical behavior and its velocity dispersion is less
than $70 Km/sec$ so it belongs to the set of galaxies that even
with modern observations can be neglected for problems of
instrumental resolution \cite{hec}. Finally our sample is
restricted to only 13 elliptical galaxies and, even if it seems
small  and derived from old data (but the sample of \cite{fer} is
smaller and the kinematical data are as old as ours) and from a
procedure "extremely sensitive to errors or incompleteness in the
data"\cite{mer}, its values of $\sigma$ give results in full
agreement with the relationship (1).
\section{The results}
As a first test we plot the data in the $(Log (\sigma/200),
LogM_{BH})$ plane and the line in Fig. 1 is obtained by using a
standard least-squares fitting and assuming, just like \cite{geb},
"that errors in dispersion measurements are zero and that errors
in $LogM_{BH}$ are the same for each galaxy"\footnote{These
hypotheses will be relaxed later.}. The best-fit line is $$Log
M_{BH} = (4.12 \pm 0.84) Log (\sigma/200) + (8.35 \pm 0.13)
\eqno(6)$$ that confirms the eq. (1) obtained by Tremaine et
al.\cite{tre}.

 The linear correlation coefficient is $r = 0.827$.
Furthermore we can calculate the unknown constant error in
$LogM_{BH}$ using the formula:
$$\epsilon_y^2 =\frac{1}{N-2} \sum_{i=1}^{N} (y_i -\alpha-\beta
x_i)^2\eqno(7)$$
 for a relation of the form $y = \alpha + \beta x$, and we obtain
 $\epsilon_y = 0.46$. If we compute the average of the experimental errors in $LogM_{BH}$ (table 1 in
 \cite{tre}),
  we obtain $<\epsilon_y>= 0.19$, so either there is an underestimation of the experimental errors
  (due for example to our hypothesis that neglects errors on $\sigma$),
  or the intrinsic dispersion of the relation is larger than the
measurement errors.

Now our aim is to study the dependence of the mass of the galaxy
inside the effective semimajor axis, on the corresponding velocity
dispersion. From the dependence of the mass-to-light ratios on the
luminosity $M/L \propto L^{0.25}$ and from the Faber - Jackson
relation $L \propto \sigma^4$, Ferrarese and Merritt \cite{fer}
infer $M \propto \sigma^5$ for early type galaxies. On the other
side, Busarello et al. \cite{busca} starting from the  data
derived in BLF (the same used in this paper) for a sample of 40
galaxies
 showed that "no real correlation holds between $M/L$ and $L$" when
the masses are computed through the Jeans equation. Then,
following the reasoning of Burkert and Silk \cite{bur}, we could
expect from the application of the Virial theorem $(M_G \propto
R_e \sigma^2)$ and from the results of the Fundamental Plane of
elliptical galaxies $(R_e \propto \sigma^2)$  that $M_G \propto
\sigma^4$, but the best-fit of our data  gives: $$Log M_{G} =
(2.16 \pm 0.71) Log (\sigma/200) + (10.98 \pm 0.11) \eqno(8)$$
This relationship, even if it has a large scatter and a poor
correlation $(r=0.678)$, induces to think that $$M_{BH} \propto
M_G \sigma^{2} \propto \sigma^{4.16} \eqno(9)$$ in agreement with
the previous fit (6).

In order  to check this hypothesis,  we show in Fig. 2 the
best-fit of the relation
$$Log(M_{BH}) = (0.87 \pm 0.17) Log \frac{M_G
\sigma^{2}}{c^2} + (4.33 \pm 0.80) \eqno(10)$$ with a correlation
coefficient $r = 0.833$ even better  (of course with our data)
than the famous relation (6).

We have also tried to consider the role of the total kinetic
energy adding the contribution due to the rotation velocity and we
obtain:
$$Log(M_{BH}) = (0.88 \pm 0.18)  Log\frac{M_G( 3\sigma^{2}+
V^2)}{c^2} + (3.80 \pm 0.92)  \eqno(11)$$ with a correlation
coefficient $r = 0.829$ and the fit is interesting too (Fig. 3).
It is remarkable that the slope of the relation (11) is the same
as the relation (10) and that the intercept changes in the way we
expect. Furthermore, we can calculate the unknown constant error
in $LogM_{BH}$ using (7) and we obtain again $\epsilon_y = 0.46$
for both the relations (10) and (11) confirming that the procedure
leads to a value 2.4 times greater than the average experimental
error. So we must check what happens if we use a fit weighted by
the experimental errors and take into account also the errors on
the independent variable.

 As for our kinematical data it is not available the
estimate of the error on $\sigma$ for each galaxy, we have
considered until now the optimistic assumption of a negligible
error.  Now it is
 time to check if something changes in the relation (10),
considering the worst scenario of a relative error $20\%$ on
$\sigma$ as we can estimate from the discussion on the possible
sources of error (anisotropy, triaxiality, fitting procedure,
inclination, etc.) contained  in BLF. To this aim we use the
effective variance method suggested by Orear \cite{or} that with
an iterative procedure minimize the $\chi^2$ given by the formula
$$\chi^2 = \sum_{i=1}^{N} \frac{(y_i -\alpha-\beta x_i)^2}{(\Delta
y_i)^2 + \beta^2 (\Delta x_i)^2}$$ for a relation of the form $y =
\alpha + \beta x$.  The corresponding results are listed in
Appendix A and show   no change in the slope of the relationships
found until now, the only advantage being  a decrease of the
difference between the residuals and the corresponding
experimental errors. On the contrary,  slightly different slopes
are obtained starting from the Akritas and Bershadi \cite{akr}
statistical method (used also in \cite{fer}) with the
corresponding $68\%$ confidence interval: $$Log(M_{BH}) = (0.98
\pm 0.09) Log \frac{M_G \sigma^{2}}{c^2} + (3.8 \pm 0.4)
\eqno(12)$$ with a reduced $\chi^2$ (see Appendix A) $\chi^2_r =
1.9$.

 {\it This fit is
shown in fig. 2 with a dashed line and involves $M_{BH} c^2
\propto M_G \sigma^2$ and could be considered as interesting as
the famous relation (1) is.}

 The same check for the relation (11) gives
$$Log(M_{BH}) = (0.995 \pm 0.082)  Log\frac{M_G(
3\sigma^{2}+ V^2)}{c^2} + (3.21 \pm 0.42)  \eqno(13)$$ (dashed
line in fig. 3) with $\chi^2_r = 1.8$.

 Even better results can be
obtained using a reduced sample (hereafter SAMPLE B) of galaxies
if we eliminate the two ellipticals with the largest residuals
N821 and N4697. The relationships obtained applying the fitting
procedure to the remaining 11 galaxies are listed in Appendix B.
The satisfying results are the increase of all correlation
coefficients, the decrease of all the values of $\chi^2_r$ and the
slopes of the two relationships that involve the kinetic energies
that are closer to the unity. While a coefficient less than one is
more difficult to be interpreted, it would be more interesting
from the theoretical point of view, the meaningful possibility of
a direct proportionality between the mass of a SMBH and the
kinetic energy of random motions of the host galaxy. Moreover,
 if our results are confirmed, the relation
between $M_{BH}$ and $M_G$ will be non linear \cite{fer2}
\cite{lao}.

 The standard scenario of  a Black Hole life
 predicts that  a part of the total mass of a SMBH is due to the
accretion process and a part of this last mass can be converted in
radiation and ejected again in the region surrounding the Hole.
From our results it seems that a part of the rest energy of the
Black Hole is in this way strictly related to the kinetic energy
of random motions of the stars of the host galaxy even in a region
far from the Hole.

\section{Conclusions}
From our sample of data we derive the suggestion to consider a
relationship between the masses of SMBHs and the Kinetic Energy of
random motions, or even the total kinetic energy, in elliptical
galaxies. Only a deeper analysis with the sophisticated machinery
today available and with new data \cite{pin}, could discriminate
if our interpretation can work and if it is generally valid or it
holds only for a restricted sample of galaxies (for example the
ellipticals with the mass in a small range of values). Given the
assumptions of the model, it is clear that triaxiality, anisotropy
of the velocity field, inclination, deviations from the $r^{1/4}$
law, are all sources of possible errors that could affect the
derived kinematical parameters. In BLF some of these errors were
discussed in detail and their total influence on the final results
was estimated in no more than $20\%$. However the underestimation
of one of the above effects, the restricted sample and even the
way to compute the masses we used, can lead to draw wrong
conclusions. On the other side we think that it would be surely
worse to neglect a priori the suggestion that comes to us by the
relationships from (10) to (13) and above all from (10B) to (13B).
So we have now arguments to ask again the question contained in
the title: is there a relationship between the mass of SMBH and
the kinetic energy in its host elliptical galaxy?
\section*{Acknowledgements}
We are grateful to Gaetano Scarpetta, Antonio D'Onofrio and Nicola
De Cesare for very useful suggestions.

\section*{Appendix A}
The formula used in this paper to estimate all the maximal errors
in the functions F of the parameters $(a,b,c,....)$ shown in table
1, is
$$\Delta F(a,b,c) = \left| \frac{\partial F}{\partial a}\right|
\Delta a + \left|\frac{\partial F}{\partial b}\right| \Delta b +
\left|\frac{\partial F}{\partial c}\right| \Delta c $$ but our
results do not change if we use for example  the formula of
Ferrarese and Merritt \cite{fer} (cited also by Tremaine et al.
\cite{tre}) to estimate the error bar in $LogM_{BH}$ that is
$$\Delta LogM_{BH}= \frac{LogM_{BH(upper)} - LogM_{BH(lower)}}{2}
$$
 In order to compare our results with the ones
obtained by \cite{fer}, we have used in (12) and (13) the same
fitting method \cite{akr} and  the reduced $\chi^2$ given by the
formula: $$\chi^2_r = \frac{1}{N-1} \sum_{i=1}^{N} \frac{(y_i
-\alpha-\beta x_i)^2}{(\Delta y_i)^2 + \beta^2 (\Delta x_i)^2}$$
for a relation of the form $y = \alpha + \beta x$.
 On the other side, the results obtained using the iterative
procedure \cite{or} (the errors $\Delta x_i$ are ignored in the
first step) are the following:
$$Log M_{BH} = (4.10 \pm 0.73) Log (\sigma/200) + (8.41 \pm 0.12) \eqno(6A)$$
with  $\chi^2_r = 1.1$;
$$Log M_{G} = (2.15 \pm 0.38) Log (\sigma/200) + (10.97 \pm 0.06) \eqno(8A)$$
with $\chi^2_r = 3.0$;
$$Log(M_{BH}) = (0.88 \pm 0.06) Log \frac{M_G
\sigma^{2}}{c^2} + (4.30 \pm 0.46) \eqno(10A)$$ with $\chi^2_r
=2.2$;
$$Log(M_{BH}) = (0.89 \pm 0.07)  Log\frac{M_G( 3\sigma^{2}+
V^2)}{c^2} + (3.77 \pm 0.52)  \eqno(11A)$$ with  $\chi^2_r = 2.1$.
\section*{Appendix B}
We neglect the two ellipticals with the largest residuals: N821
and N4697. The relationships contained in the main part of the
paper are reanalyzed  using the remaining 11 galaxies forming the
so called SAMPLE B. The results obtained using the iterative
procedure \cite{or} are the following:
$$Log M_{BH} = (4.06 \pm 0.73) Log (\sigma/200) + (8.46 \pm 0.13) \eqno(6B)$$
with a linear correlation coefficient $r = 0.846$ and  $\chi^2_r =
0.9$;
$$Log M_{G} = (2.26 \pm 0.52) Log (\sigma/200) + (10.90 \pm 0.09) \eqno(8B)$$
with $r = 0.755$ and $\chi^2_r = 1.6$;
$$Log(M_{BH}) = (0.92 \pm 0.07) Log \frac{M_G
\sigma^{2}}{c^2} + (4.25 \pm 0.49) \eqno(10B)$$ with  $r = 0.907$
and $\chi^2_r = 0.9$;
$$Log(M_{BH}) = (0.92 \pm 0.07)  Log\frac{M_G( 3\sigma^{2}+
V^2)}{c^2} + (3.70 \pm 0.55)  \eqno(11B)$$ with a $r = 0.907$ and
$\chi^2_r = 0.9$.

 The results obtained using the Akritas and
Bershadi method \cite{akr} are the following:
$$Log(M_{BH}) = (1.03 \pm 0.08) Log \frac{M_G
\sigma^{2}}{c^2} + (3.71 \pm 0.38) \eqno(12B)$$ with $\chi^2_r =
1.06$;
$$Log(M_{BH}) = (1.06 \pm 0.08)  Log\frac{M_G( 3\sigma^{2}+
V^2)}{c^2} + (3.02 \pm 0.39)  \eqno(13B)$$ with  $\chi^2_r =
1.01$.
\newpage
\pagestyle{empty}
\begin{figure}
  \centering
\resizebox{\hsize}{!}{\includegraphics{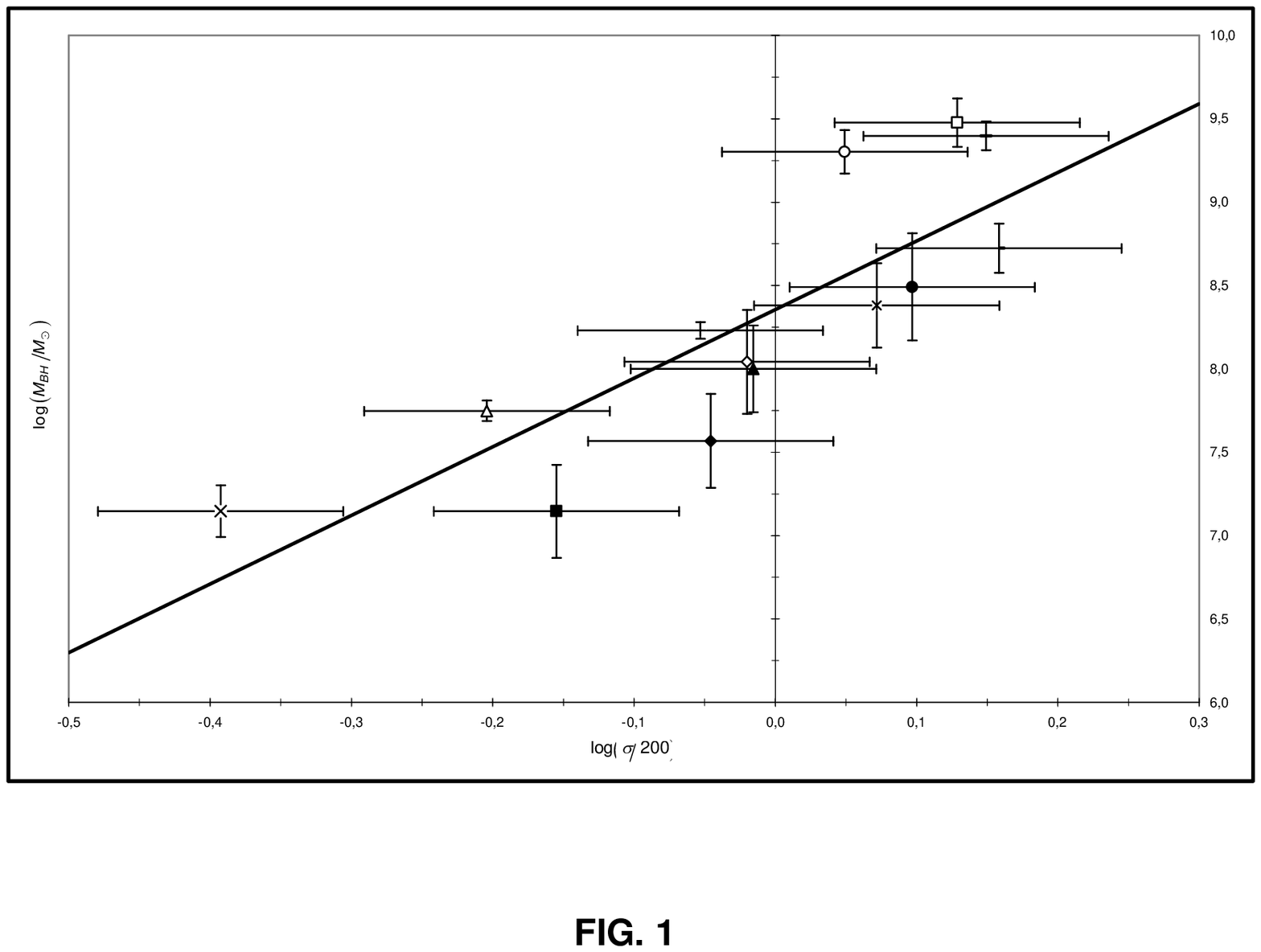}} \caption{
SMBH mass versus luminosity weighted, line of sight velocity
dispersion of the host elliptical galaxy. The solid line is the
best linear fit with the least squares method (6). The error bars
for $M_{BH}$ are from \cite{tre} and for $\sigma$ are calculated
from the upper limit on the relative error of $20\%$ estimated
from BLF.} \label{Fig. 1}
\end{figure}
\begin{figure}
  \centering
\resizebox{\hsize}{!}{\includegraphics{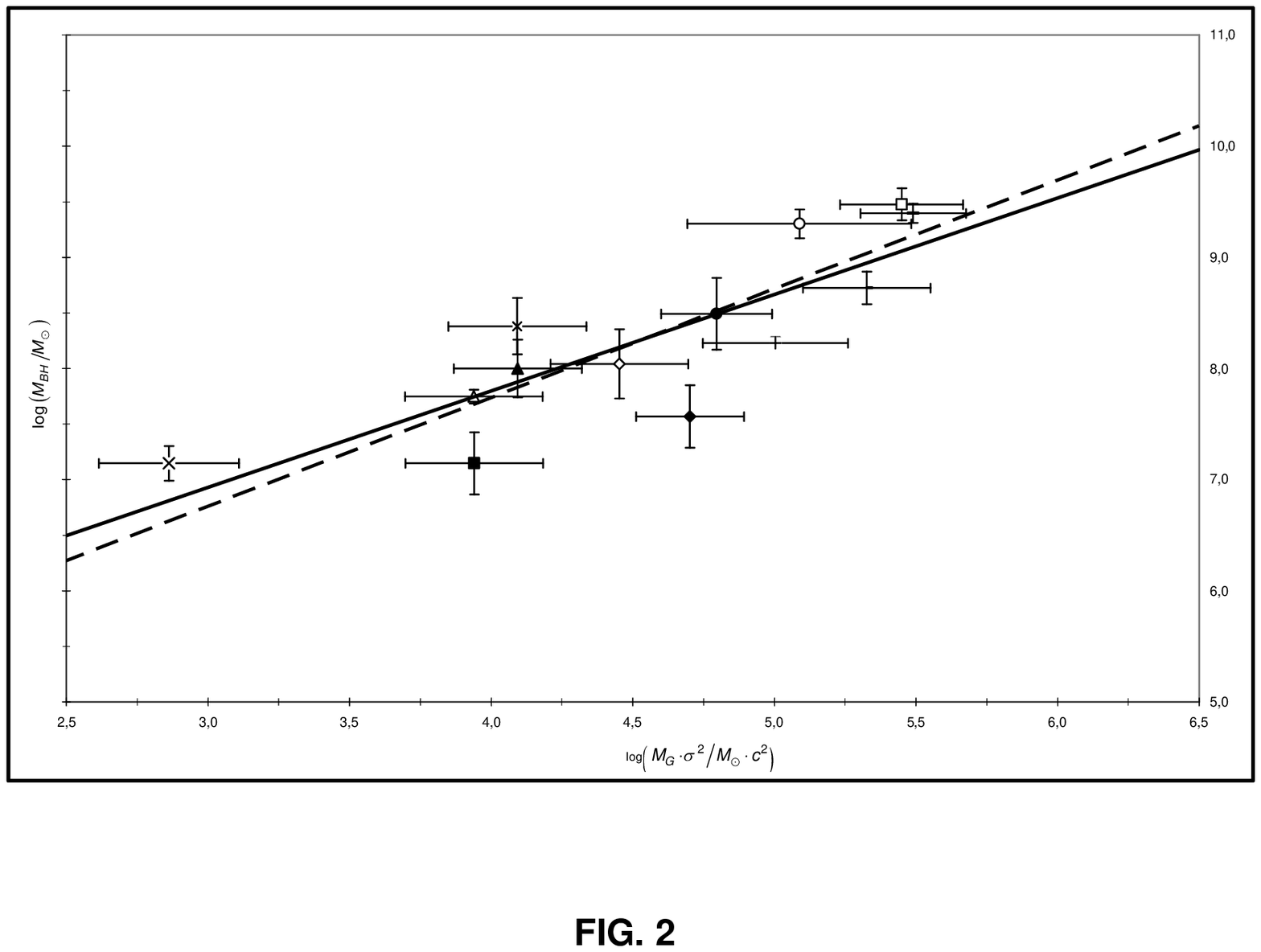}} \caption{
SMBH mass versus $M_G \sigma^2/c^2$ of the host elliptical galaxy.
The solid line is the best linear fit (10) with the least squares
method, the dashed line is the best fit (12) with the Akritas -
Bershadi method \cite{akr}. The error bars for $M_{BH}$ are from
\cite{tre}, for $M_G$ are in table 1 and for $\sigma$ are
calculated from the upper limit on the relative error of $20\%$
estimated from BLF.} \label{Fig. 2}
\end{figure}
\begin{figure}
  \centering
\resizebox{\hsize}{!}{\includegraphics{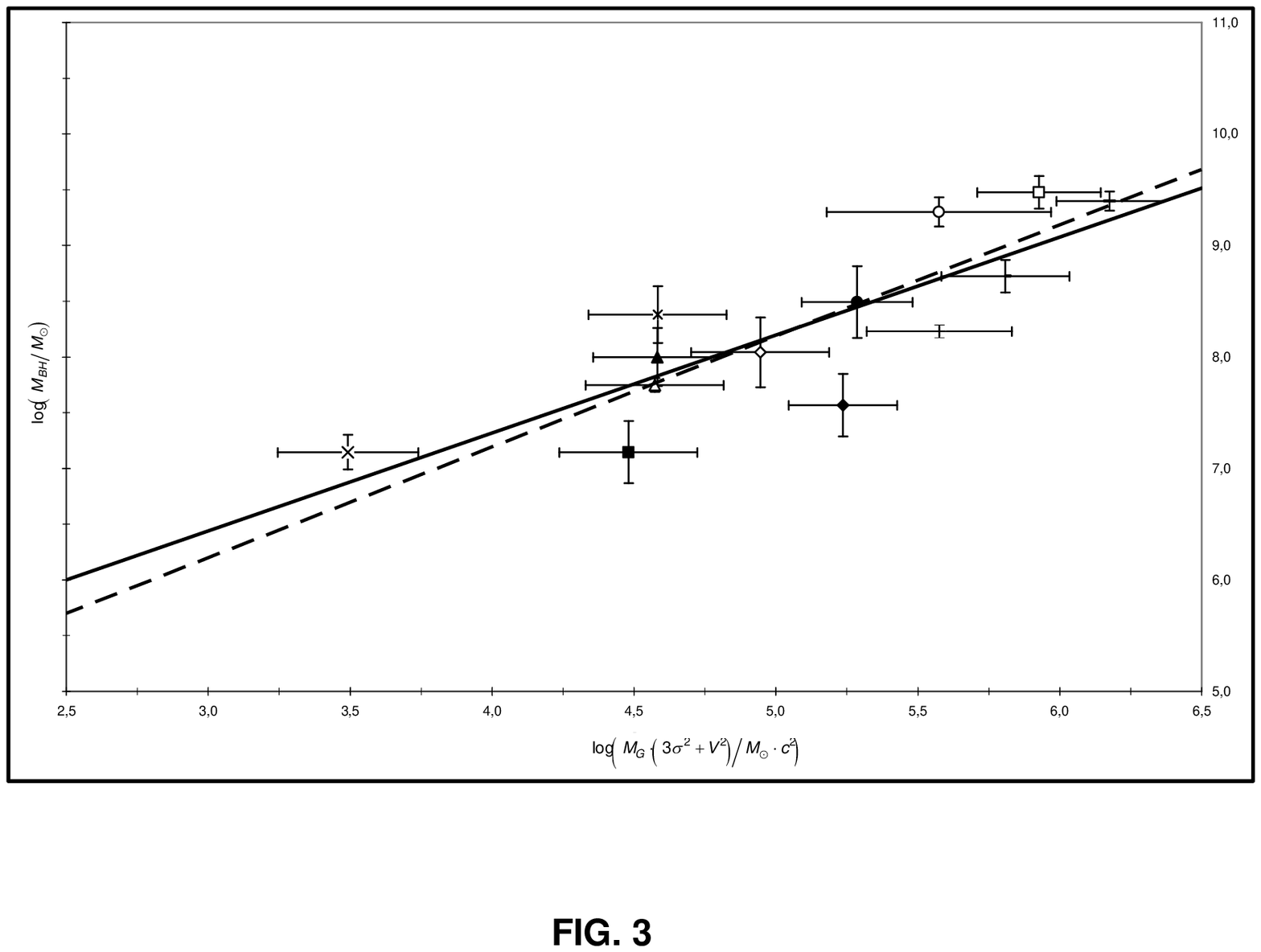}}
\caption{SMBH mass versus $M_G (3\sigma^2+ V^2)/c^2$ of the host
elliptical galaxy. The solid line is the best linear fit (11) with
the least squares method, the dashed line is the best fit (13)
with the Akritas - Bershadi method \cite{akr}. The error bars for
$M_{BH}$ are from \cite{tre}, for $M_G$ are in table 1 and for
$\sigma$ and $V$ are calculated from the upper limit on the
relative error of $20\%$ estimated from BLF.} \label{Fig. 3}
\end{figure}
\end{document}